\documentclass[preprint,
superscriptaddress,
amsmath,amssymb,aps,showkeys,showpacs,
twoside,final,secnumarabic,
nofootinbib]{revtex4-2}

\usepackage[paperwidth=205mm,paperheight=290mm,top=17mm,bottom=25mm,
inner=17mm,outer=17mm,
twoside]{geometry}

\usepackage{cmap} 
\usepackage[T1,T2A]{fontenc}
\usepackage[utf8]{inputenc}
\usepackage[russian,english]{babel}
\usepackage{color}
\usepackage{graphicx}
\usepackage{dcolumn}
\usepackage{bm} 
\usepackage[unicode=true,colorlinks=true,linkcolor=magenta, urlcolor=blue, citecolor = blue,breaklinks]{hyperref}
\usepackage{multirow}
\usepackage{url}
\usepackage{breakurl}
\DeclareGraphicsExtensions{.eps}

\newcount\issue
\newcount\Vol
\newcount\numb
\usepackage{fancyhdr} 
\def\Vol{\textbf{80}}
\def\numb{x}
\begin{document}

\title{ CONFERENCE SECTION \\[20pt]
Applying Normalizing Flows for spin correlations reconstruction in associated top-quark pair and dark matter production} 

\def\addressa{Skobeltsyn Institute of Nuclear Physics of Lomonosov Moscow State University (SINP MSU), 1(2), Leninskie gory, GSP-1, Moscow 119991, Russian Federation}

\author{\firstname{E.}~\surname{Abasov}}
\email[E-mail: ]{emil@abasov.ru }
\affiliation{\addressa}
\author{\firstname{L.}~\surname{Dudko}}
\affiliation{\addressa}
\author{\firstname{E.}~\surname{Iudin}}
\affiliation{\addressa}
\author{\firstname{A.}~\surname{Markina}}
\affiliation{\addressa}
\author{\firstname{P.}~\surname{Volkov}}
\affiliation{\addressa}
\author{\firstname{G.}~\surname{Vorotnikov}}
\affiliation{\addressa}
\author{\firstname{M.}~\surname{Perfilov}}
\affiliation{\addressa}
\author{\firstname{A.}~\surname{Zaborenko}}
\affiliation{\addressa}
\received{xx.xx.2025}
\revised{xx.xx.2025}
\accepted{xx.xx.2025}

\begin{abstract}
We apply a unified machine‐learning framework based on Normalizing Flows (NFs) for the event‐by‐event reconstruction of invisible momenta and the subsequent evaluation of spin‐sensitive observables in top‐quark pair and dark‐matter (DM) associated production processes. Building on recent studies in single‐top + DM topologies, we extend the research to $t\bar{t} + DM$ final state. Inputs to our networks combine low‐level four‐momenta and missing transverse energy with high‐level kinematic and angular variables. We compare a baseline multilayer perceptron (MLP) regressor, an autoregressive flow and the conditional ‘‘$\nu$‐Flows’’ model—trained to learn the full conditional density. In these final states all the models perform well and demonstrate high reconstruction quality in independent regions split by $m_{t\bar{t}}$ for validation purposes. We highlight the potential of this approach to be extended to 3 and 4 top-quark production.
\end{abstract}

\par
\keywords{Dark matter; top-quark; angular correlations   \\[5pt]}

\maketitle
\thispagestyle{fancy}


\section*{Introduction}
The top quark, owing to its large mass and extremely short lifetime, preserves its spin information until decay. This unique property makes top-quark observables based on angular correlations a powerful probe of the Standard Model (SM) and of physics beyond it. Spin-sensitive measurements have already played a crucial role in precision studies of top-quark pair production and decay. The spin correlations between top and antitop quarks in pair production have also been used to probe the effects of quantum entanglement~\cite{ATLAS:2023fsd, CMS:2024pts}, and they are also highly promising in new physics scenarios.

A particularly instructive example arises in collider searches for dark matter (DM) in simplified models with light scalar or pseudoscalar mediators that couple preferentially to third-generation quarks, in line with minimal flavor violation~\cite{Chivukula1987CompositetechnicolorSM,PhysRevLett.65.2939,BURAS2001161,DAMBROSIO2002155,PhysRevD.88.063510}. In this framework, associated production of DM with top quarks ($t\bar{t}+\mathrm{DM}$ or $t/\bar{t}+\mathrm{DM}$) has been actively explored by ATLAS and CMS~\cite{CMS:2019zzl,ATLAS:2022ygn}, motivated more broadly by robust astrophysical and cosmological evidence for DM~\cite{Rubin:1970zza,Harvey:2015hha}. In this context, new spin-sensitive angular variables can significantly enhance the separation of signal from SM backgrounds. However, their practical use faces a key challenge: the presence of one or more invisible particles (such as neutrinos and potential DM mediators) complicates the reconstruction of the top-quark rest frame and, consequently, the accurate evaluation of spin-correlation observables.

Recent work~\cite{Abasov:2024nec} introduced an DM sensitive angular variable based on the cosine of the angle between the charged lepton and the down-type quark in the top-quark rest frame. This observable is informative only when the top rest frame is reconstructed using all invisible contributions, i.e., both the neutrino and the DM mediator, rather than only the neutrino. Achieving such an event-by-event separation of invisible momenta is nontrivial. Analytical strategies were explored~\cite{Abasov:2024nec,Bunichev:2024} but did not yield the required accuracy in realistic conditions, particularly once detector effects are included.

To address this, machine-learning methods have been deployed to reconstruct the 4-momenta of invisible particles and thus restore the spin-sensitive top-quark kinematics. Two complementary families of approaches have been investigated: a baseline feed-forward neural network (multilayer perceptron, MLP) and flow-based generative models known as Normalizing Flows (NFs)~\cite{Kobyzev_2021}. The latter have been found to be particularly well-suited to collider problems with invisible final states because they learn the full conditional multi-dimensional probability density of the unobserved degrees of freedom given the observed event. This makes them powerful tools for preserving physical correlations, including spin information.

The aim of the present article is to generalize this strategy to pair top-quark production and provide an insight into future research on 3 and 4 top-quark systems, where spin correlations offer a wealth of information but event reconstruction is substantially more complex. In $t\bar{t}$ and especially in $t\bar{t}t\bar{t}$ final states, multiple neutrinos and combinatorics associated with the proper top-quark reconstruction blur the direct connection between measured objects and underlying parton-level kinematics. We show that the flow-based reconstruction paradigm scales naturally to these multi-top systems and enables precise recovery of spin-sensitive angular observables, paving the way for robust measurements and for enhanced sensitivity to BSM effects that couple to the top sector.

\section{Reconstructing spin-sensitive observables with machine learning}
We briefly summarize the methodology validated in associated single-top and scalar DM mediator studies~\cite{Abasov:2024nec}, which we adapt and extend here to two- and four-top quark systems.
We use a Simplified Model with scalar DM mediator $\Phi$, which is the most difficult state to separate from SM contribution:
\begin{equation}
L_{\Phi}=g_\chi \Phi \bar{\chi} \chi+\frac{g_v \Phi}{\sqrt{2}} \sum_f\left(y_f \bar{f} f\right) 
\end{equation}
Here, $\chi$ is a fermionic dark matter particle, $f$ - SM fermions.
Event samples are generated for proton-proton collisions at 13 TeV using CompHEP~\cite{CompHEP:2004qpa,Pukhov:1999gg} and MadGraph5\_aMC@NLO~\cite{Alwall:2014hca} with NNPDF23\_nlo\_as\_0118 parton distribution functions~\cite{Buckley:2014ana}. Following the LHC DM Working Group guidance~\cite{Boveia:2016mrp}, representative DM simplified model benchmarks are adopted, e.g., $g_f=g_\chi=1$, $m_\chi=1$ GeV, and $m_\Phi=400$ GeV (consistent with current collider limits~\cite{CMS:2021eha}). Parton-level events are hadronized in Pythia8~\cite{Bierlich:2022pfr} and passed through a fast detector simulation with DELPHES~\cite{deFavereau:2013fsa}. Analyses are performed both at generator level and after detector smearing to assess realism and the need for unfolding.

The learning target is the set of invisible 4-momenta in each event: these are neutrino and the mediator in the single-top production used in the previous article and neutrinos for the $t\bar{t}$ production. Inputs to the networks include:
\begin{itemize}
    \item Low-level features: four-momenta of reconstructed objects and missing transverse momentum (MET); absent objects are zero-padded for fixed-size inputs.
    \item High-level features: physically motivated combinations such as invariant masses, scalar products of four-momenta, and intermediate resonance reconstructions (e.g., $W$, top candidates).
\end{itemize}

All input and target variables are standardized. Datasets are split into train/validation/test partitions. For MLP, which is used as a baseline in this work, the objective function is mean absolute error (MAE) on invisible momenta. Because the physics goal is the fidelity of the reconstructed angular distribution, we additionally evaluate histogram-level metrics on spin-sensitive observables, following~\cite{Abasov:2024nec}, such as a histogram MAE and a $\chi^2$ score comparing reconstructed and truth distributions.

Normalizing Flows differ from deterministic regressors by learning a conditional probability density $p(\text{invisible}|\text{visible})$. In practice, we employ the \texttt{nflows}~\cite{conor_durkan_2020_4296287} and PyTorch~\cite{Ansel_PyTorch_2_Faster_2024} packages and consider two variants of this neural network type:
\begin{itemize}
    \item A basic autoregressive NF, primarily as a baseline generative model (Basic Flows).
    \item The $\nu$-Flows architecture~\cite{Leigh:2022lpn}, which stacks coupling layers with piecewise rational-quadratic spline transforms, conditioned on a compact learned representation of the visible event. This design has proven the most effective in single-top topologies and is thought to be naturally extensible to more complex final states.
\end{itemize}

For the encoder networks providing the conditional context, compact fully connected architectures are used, with hyperparameters tuned on validation data. At inference time, multiple NF samples per event are drawn and aggregated (e.g., via the median) to obtain robust point estimates of the invisible momenta. The resulting four-vectors are then used to reconstruct the relevant top-quark rest frames and compute angular observables.

In the single-top plus mediator study~\cite{Abasov:2024nec}, this strategy demonstrated that NF-based models deliver a substantial improvement over MLPs in reproducing the target angular distributions, both at parton level and after detector smearing. While MLPs can reach smaller per-component MAE on momenta, they degrade the shape fidelity of the final spin observable; in contrast, $\nu$-Flows achieve an order-of-magnitude better agreement in histogram-based metrics (e.g., $\chi^2$) for the discriminating cosine observable defined in the top rest frame constructed with all invisible contributions. This confirms that modeling the full conditional density is crucial when the physics target depends sensitively on multi-dimensional correlations.

\section{From single-top to $t\bar{t}$ spin correlations}
The extension to $t\bar{t}$ final state is compelling, since the spin correlations in $t\bar{t}$ are well-established precision observables, and have been successfully used to measure quantum entanglement in top-antitop pair. Their measurement also benefits from accurate per-event assignment and reconstruction, especially in dileptonic channels with two neutrinos.

In this case, the central technical obstacle mirrors the single-top plus mediator case: multiple invisible particles must be separated to reconstruct parent rest frames and to retain spin information encoded in angular patterns. We use the entanglement marker $D= Tr[C]/3$ introduced in~\cite{Afik:2020onf}, which arises from spin density matrix decomposition $$
\rho \propto \tilde{A} I \otimes I+\tilde{B}_i^{+} \sigma^i \otimes I+\tilde{B}_i^{-} I \otimes \sigma^i+\tilde{C}_{i j} \sigma^i \otimes \sigma^j
$$ and can be experimentally measured from differential cross-section $$\frac{1}{\sigma} \frac{\mathrm{~d} \sigma}{\mathrm{~d} \cos \varphi}=\frac{1}{2}(1-D \cos \varphi)$$ Here $\varphi$ is the angle between the two lepton directions measured in their parent top quark and antiquark rest frames. Additionally, we measure the same variable in the Simplified DM model with scalar mediator in $t\bar{t}\Phi$ signature, using only neutrinos to reconstruct top-quark rest frames.

\begin{figure}[ht]
    \centering
    \includegraphics[width=0.6\linewidth]{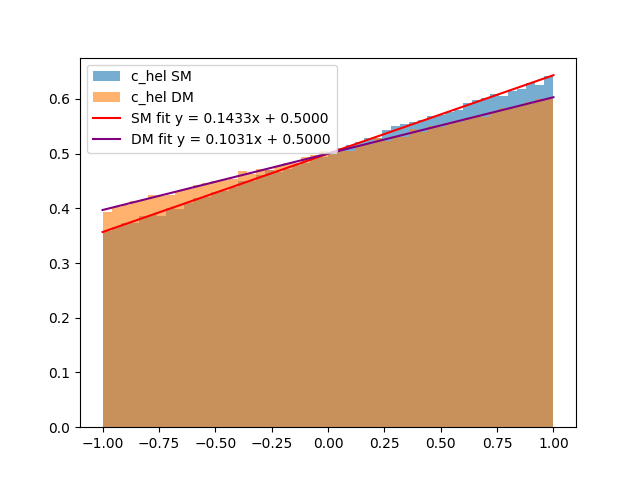}
    \caption{Entanglement marker $cos\varphi$ in parton-level samples for $t\bar{t}$ (SM) and $t\bar{t}\Phi$ (DM) signatures, whole dataset is used.}
    \label{fig:2top_c_hel}
\end{figure}

\begin{figure}[ht]
    \centering
    \includegraphics[width=0.9\linewidth]{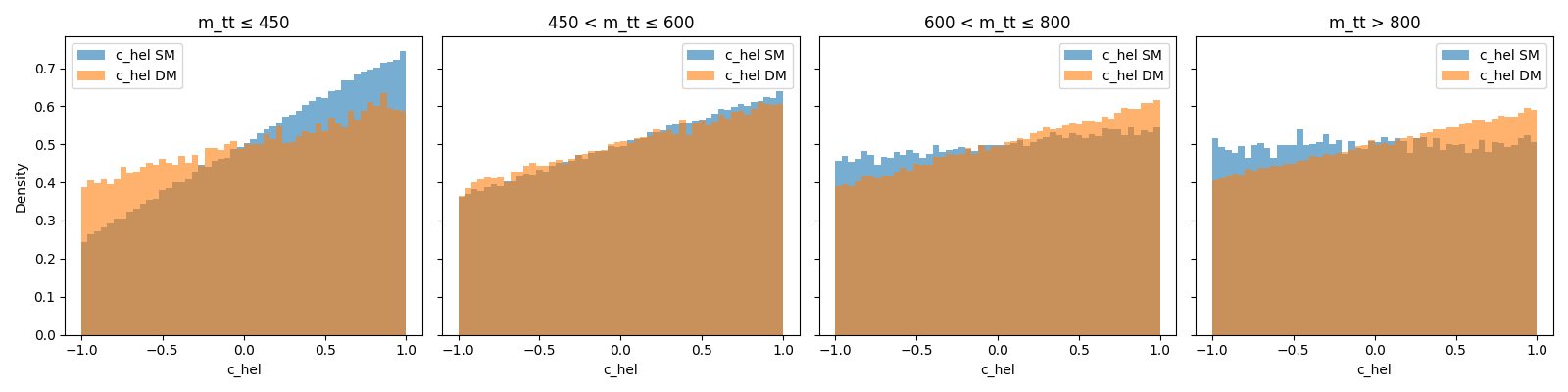}
    \caption{Entanglement marker $cos\varphi$ in parton-level samples for $t\bar{t}$ (SM) and $t\bar{t}\Phi$ (DM) signatures, data is split into distinct $m_{t\bar{t}}$ regions.}
    \label{fig:2top_c_hel_split}
\end{figure}

Here, final state leptons can be matched to top-quark resonances using their charge, so the ground truth values for all the variables are known with high precision. We measure $cos\varphi$ (also denoted as $c_{hel}$ in the plots) both for the whole data sample on Fig.~\ref{fig:2top_c_hel}, as well as the split by $m_{t\bar{t}}$ into different kinematic regions (Fig.~\ref{fig:2top_c_hel_split}). It can be seen that both SM and DM samples behave differently depending on the $m_{t\bar{t}}$, which is used to provide independent validation for our machine learning methods - training is conducted on the whole dataset, while final results are provided on the $m_{t\bar{t}}$ splits.


Flow-based conditional densities offer a unified way to tackle these topologies. By learning $p(\text{invisible}|\text{visible})$ with all kinematic and combinatorial constraints implicitly included, NFs preserve the high-dimensional structure linking observed objects, MET, and the invisible momenta across the entire event. Once trained, they provide fast, flexible samplers for event-by-event reconstruction, which can be combined with standard unfolding techniques to correct for detector effects and compare directly to parton-level theory predictions.

\subsection{Application of machine learning reconstruction}
We follow the framework of~\cite{Abasov:2024nec} in order to facilitate the creation of an universal network tailored to reconstructing spin correlations in processes with top-quark. Charged particles 4-momenta as well as ``high-level'' energy-based and spin-based variables are used as the network input, neutrinos and mediator 4-momenta are the outputs. The data is padded with zeroes to allow the network to work with SM samples, which lack the mediator. All data is normalized and 1/1 mix of DM and SM samples is used to train the network.

\begin{figure}[ht]
    \centering
    \includegraphics[width=0.3\linewidth]{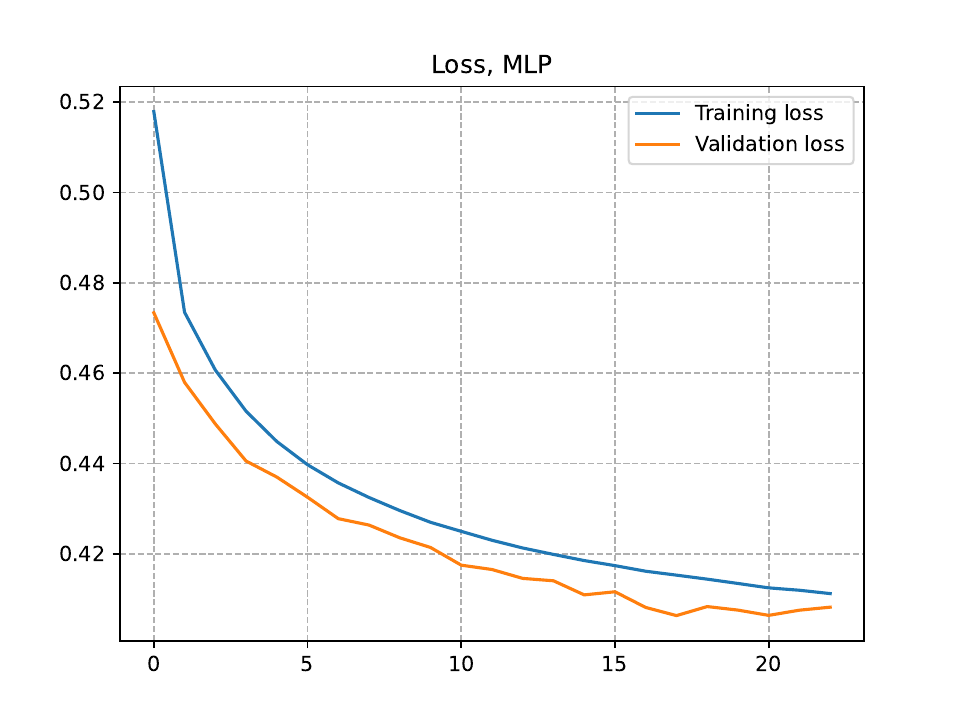}
    \includegraphics[width=0.3\linewidth]{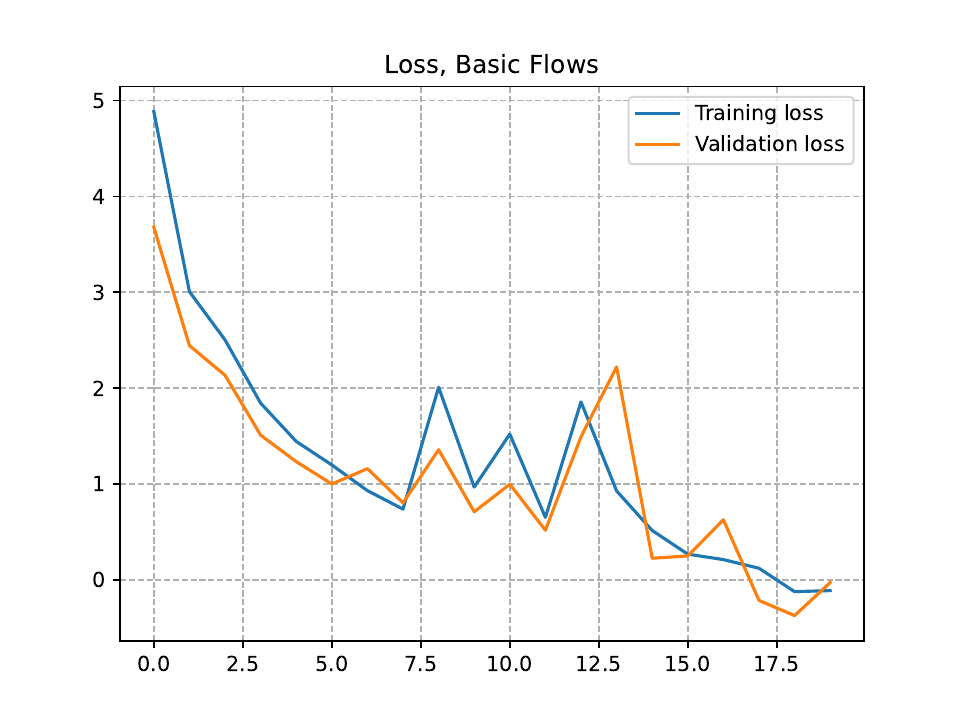}
    \includegraphics[width=0.3\linewidth]{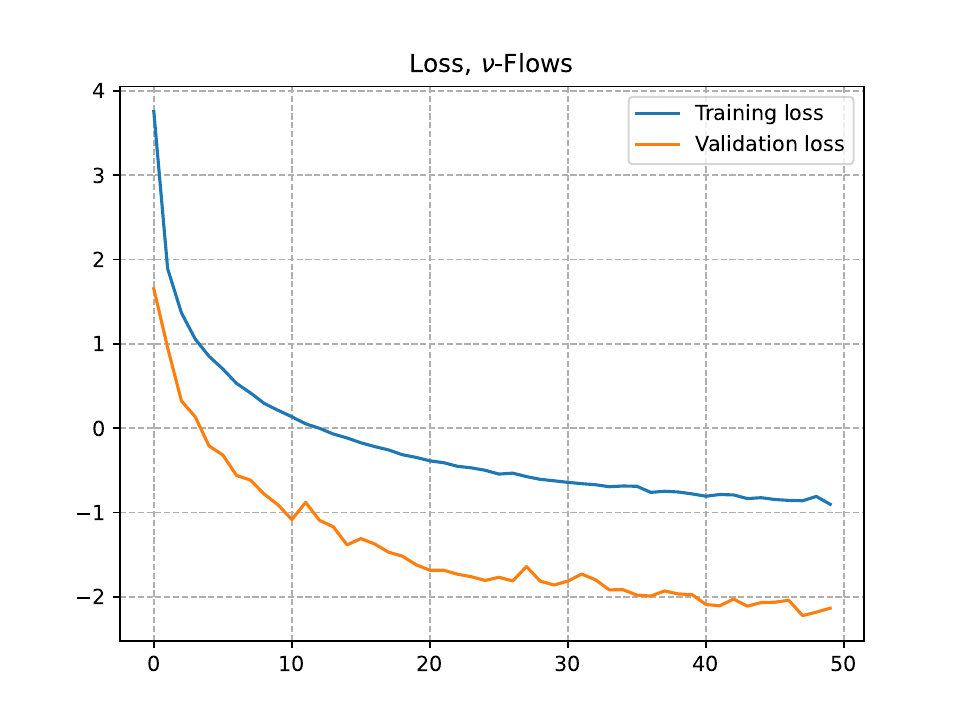}
    \caption{Loss functions of the neural networks by epoch. L1 loss is used to train the MLP, NF-based architectures use the likelihood.}
    \label{fig:loss}
\end{figure}

Fig.~\ref{fig:loss} depicts loss functions of all the networks used in this articles. MLP is trained using the L1 loss, while both Basic Flows and $\nu$-Flows utilize the likelihood.

\begin{figure}[ht]
    \centering
    \includegraphics[width=0.45\linewidth]{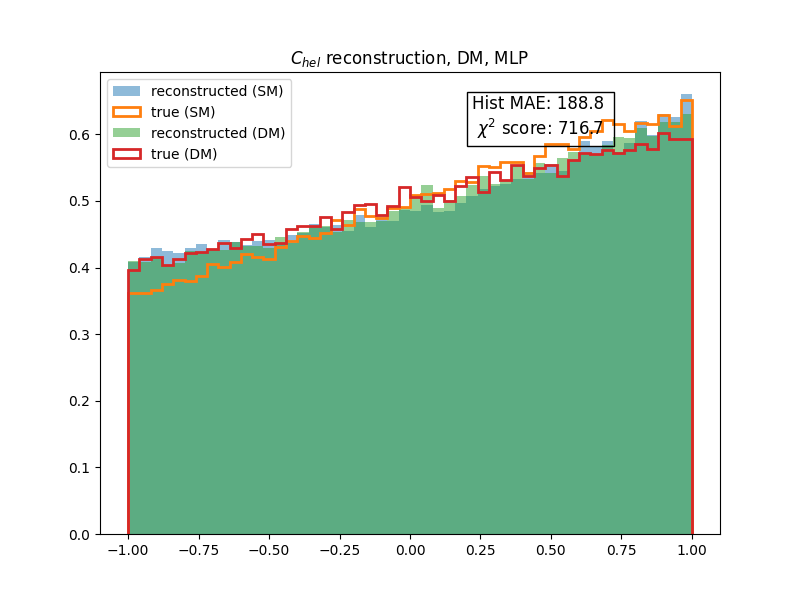}
    \includegraphics[width=0.45\linewidth]{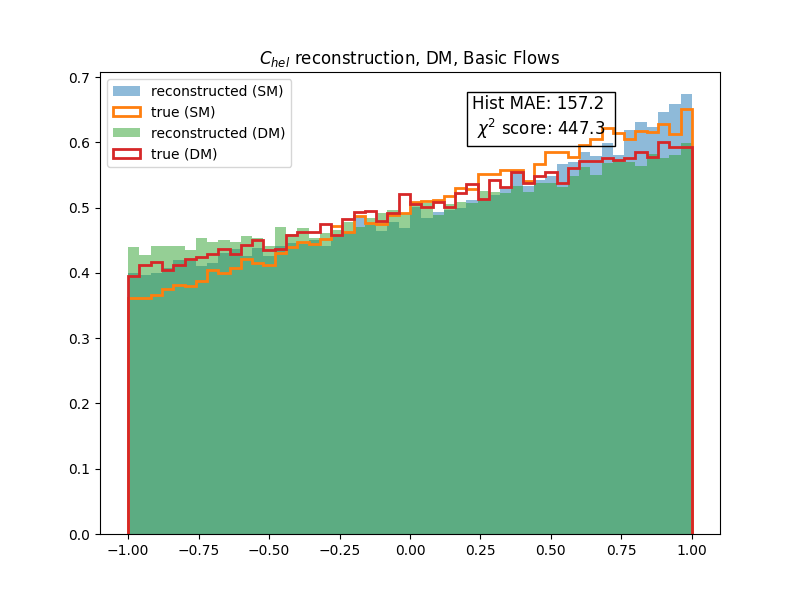}
    \includegraphics[width=0.45\linewidth]{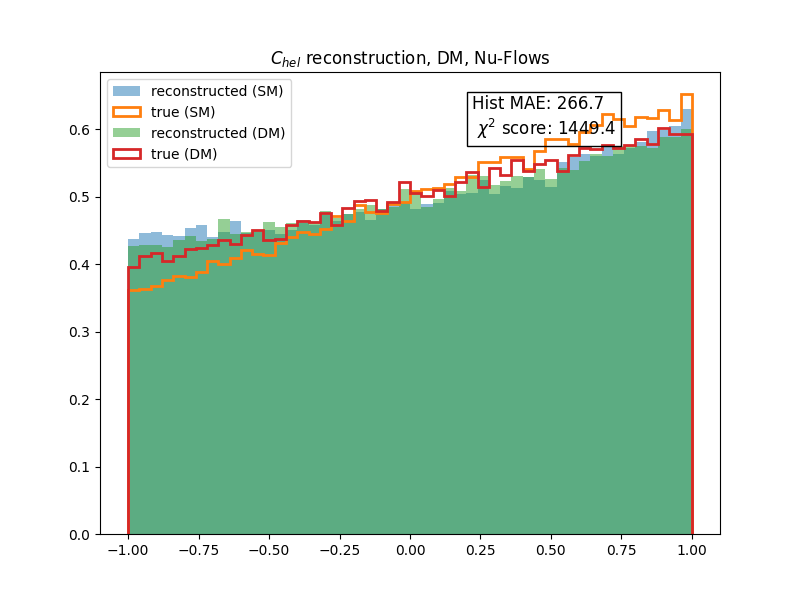}
    \caption{Reconstructed $\cos\varphi$ for both SM and DM samples and 3 types of neural networks - MLP, Basic Flows and $\nu$-Flows.}
    \label{fig:c_hel_pred}
\end{figure}

\begin{figure}[ht]
    \centering
    \includegraphics[width=0.8\linewidth]{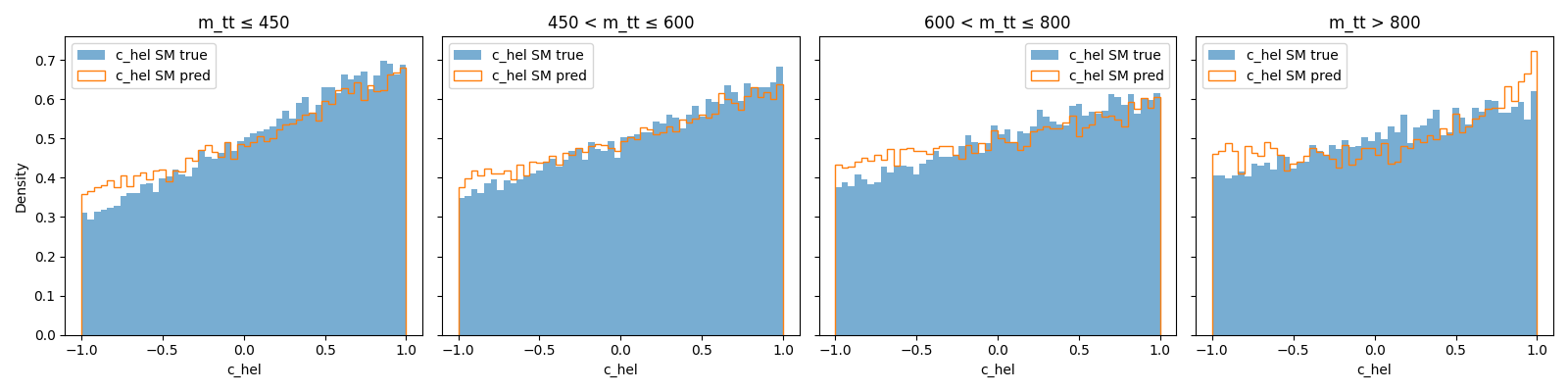}
    \includegraphics[width=0.8\linewidth]{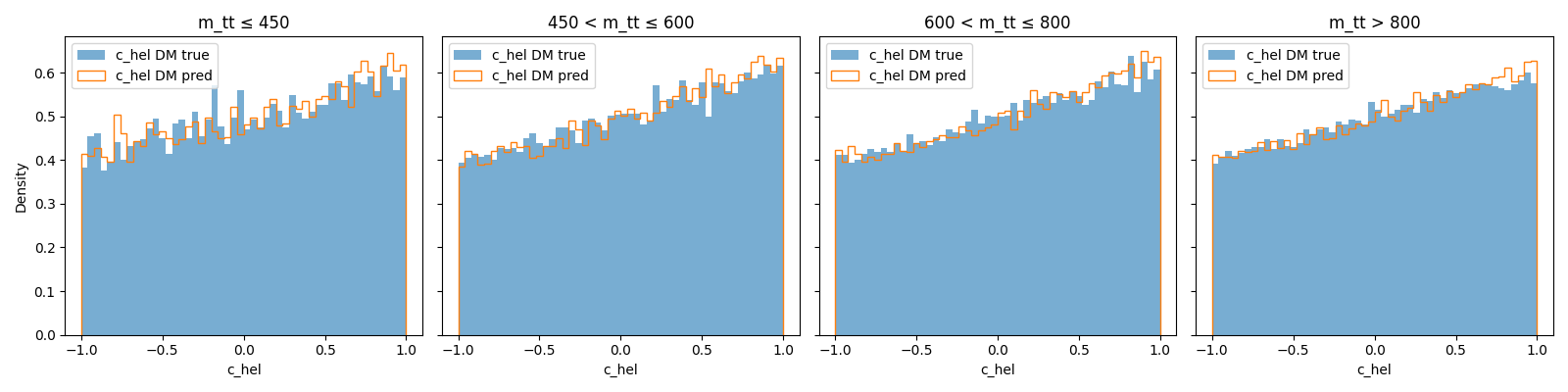}
    \caption{Reconstructed $\cos\varphi$ for both SM and DM samples with $m_{t\bar{t}}$ splits for the Basic Flows.}
    \label{fig:c_hel_pred_mlp}
\end{figure}

\begin{figure}[ht]
    \centering
    \includegraphics[width=0.8\linewidth]{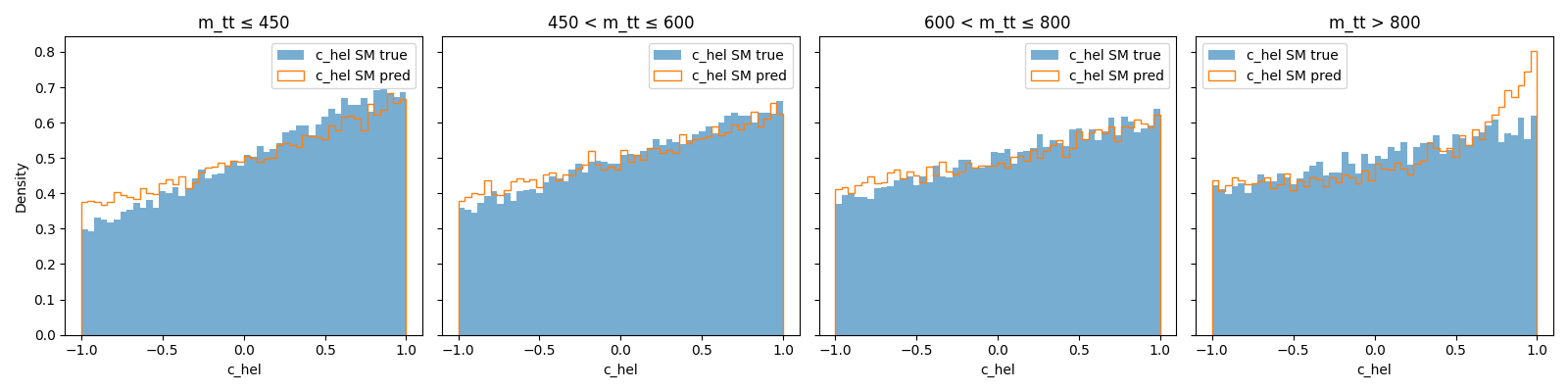}
    \includegraphics[width=0.8\linewidth]{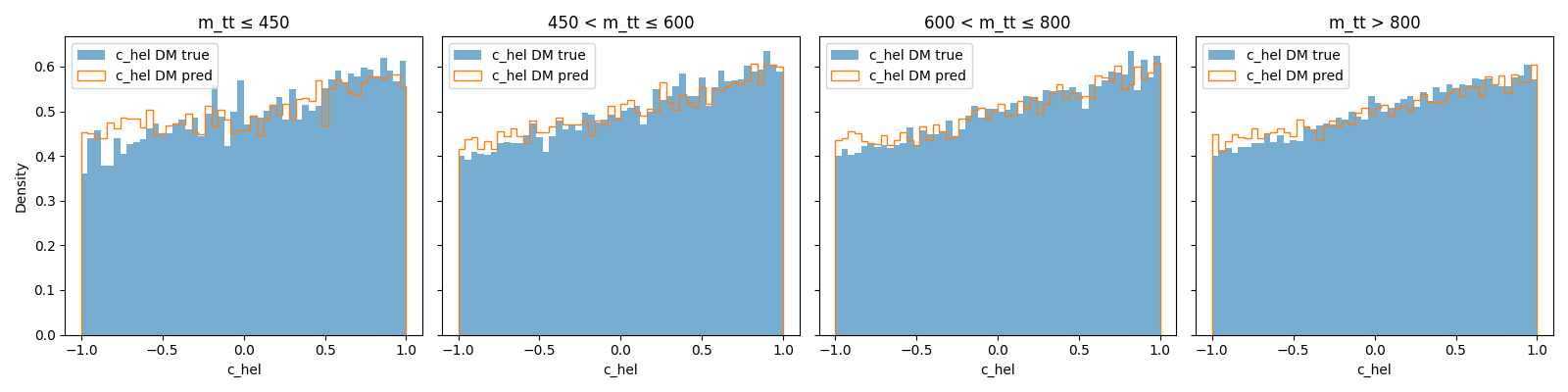}
    \caption{Reconstructed $\cos\varphi$ for both SM and DM samples with $m_{t\bar{t}}$ splits for the MLP.}
    \label{fig:c_hel_pred_basic}
\end{figure}

\begin{figure}[ht]
    \centering
    \includegraphics[width=0.8\linewidth]{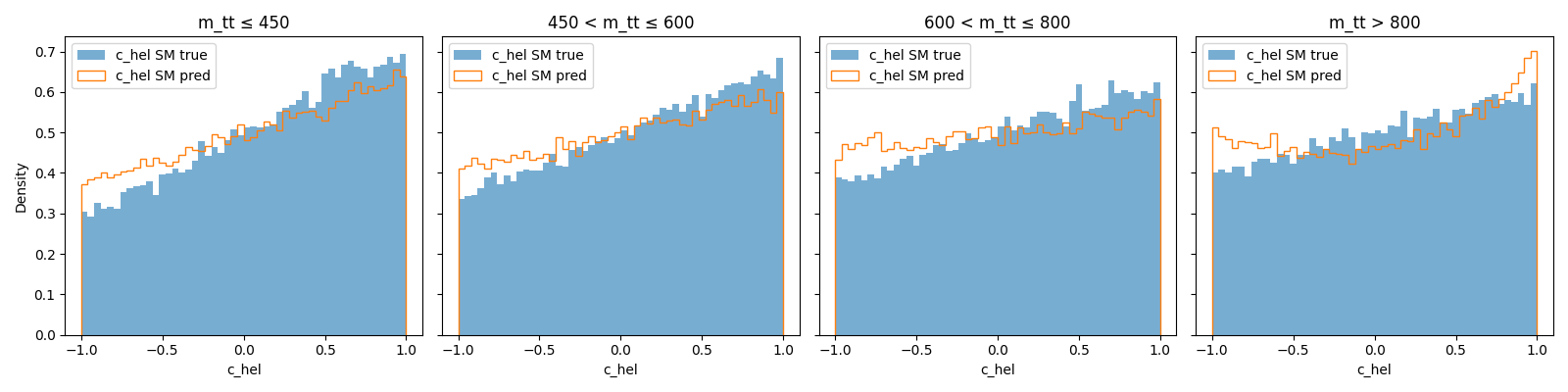}
    \includegraphics[width=0.8\linewidth]{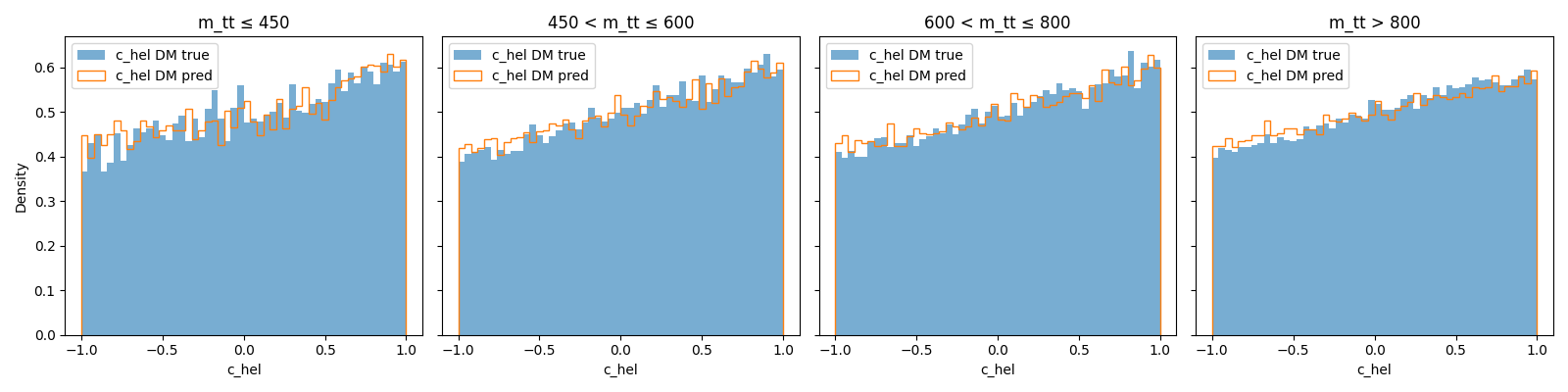}
    \caption{Reconstructed $\cos\varphi$ for both SM and DM samples with $m_{t\bar{t}}$ splits for the $\nu$-Flows.}
    \label{fig:c_hel_pred_nuflows}
\end{figure}

After training, all 3 networks were evaluated on both the SM and DM datasets separately. The $\cos\varphi$ distribution for these samples is shown on Fig.~\ref{fig:c_hel_pred}, $m_{t\bar{t}}$ splits are depicted for the MLP, Basic Flows and $\nu$-Flows are presented on Fig.~\ref{fig:c_hel_pred_mlp},~\ref{fig:c_hel_pred_basic},~\ref{fig:c_hel_pred_nuflows}, respectively.


As can be seen, all architectures achieve an acceptable level of agreement on the target variable in all $m_{t\bar{t}}$ regions. It should be stressed, that the $m_{t\bar{t}}$ split is done only for the inference and is not used in the training process in any way.

\section{Prospects: 3 and 4 top-quark production} 
\begin{figure}[!h!]
\centering
\includegraphics[width = .6\linewidth]{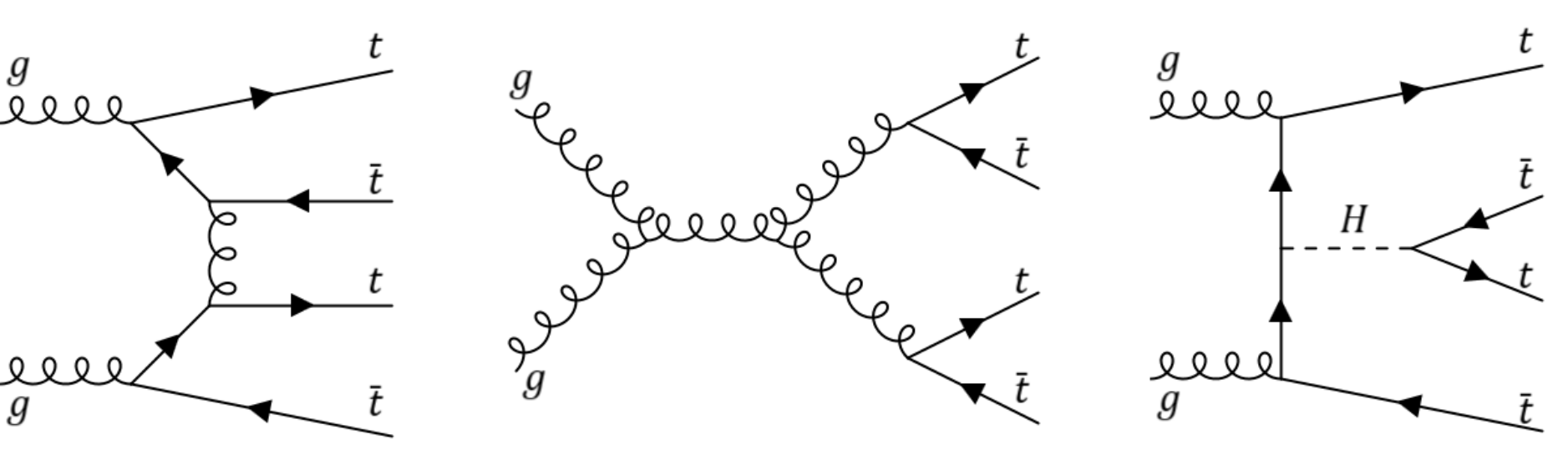}
\includegraphics[width = .6\linewidth]{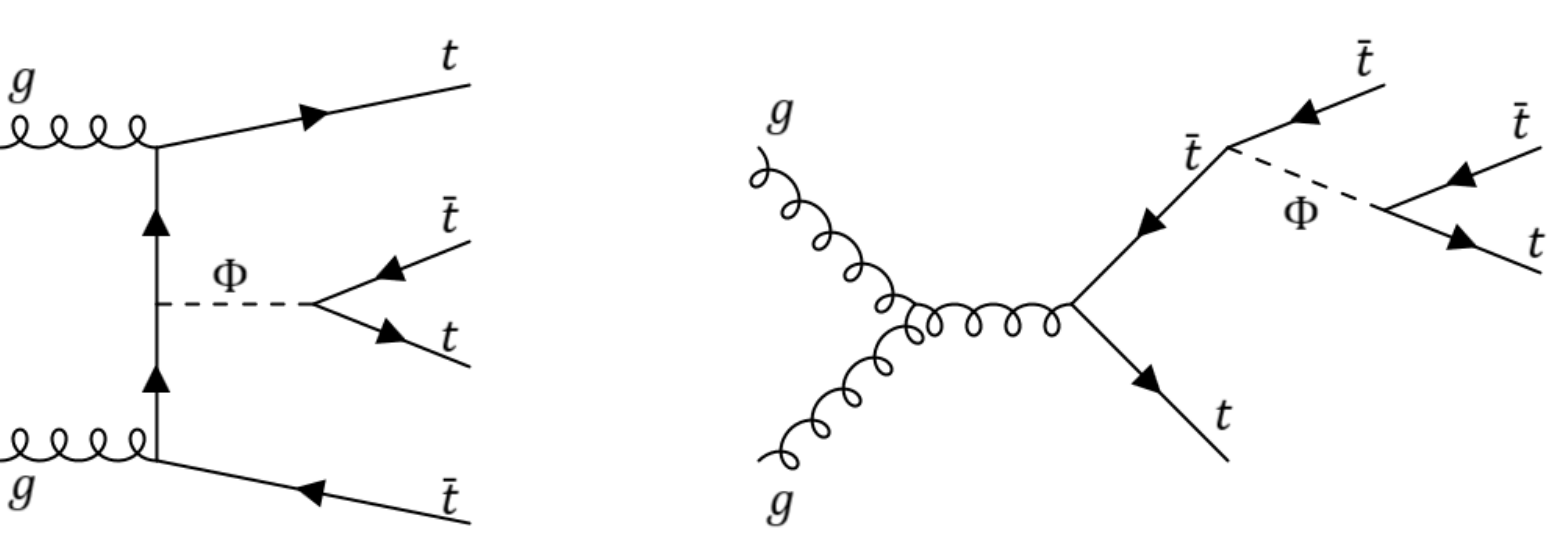}
\caption{Representative Feynman diagrams describing the production of four top quarks in the collision of two gluons from two colliding hadrons within the Simplified Dark Matter Model with a scalar mediator.}
\label{fig:4t_diagrams}
\end{figure}

Though $t/\bar{t}+\Phi, t\bar{t}+\Phi$ processes have been the most common in the search for dark matter in association with top-quarks, rare processes of $t\bar{t}tW$ and $t\bar{t}t\bar{t}$ production were shown to be very sensitive to the presence of dark matter in the framework of the Simplified Models due to the $\Phi \to t\bar{t}$ decays~\cite{Abasov:2024mwk}. Inclusion of the variables sensitive to top-quark spin correlations can prove vital to the future searches for dark matter in these states. As can be seen from the Feynman diagrams of the $t\bar{t}t\bar{t}$ process (Fig.~\ref{fig:4t_diagrams}), the approach discussed in this article for the $t\bar{t}$ production can be directly applied in this process for the pairs originating in $gt\bar{t}$ and $\Phi t\bar{t}$ vertices. However, the task of selecting these pairs is complicated due to the presence of combinatorial ambiguities and requires additional studies, such as application of transformer-based foundation models for unified unfolding and resonance reconstruction.
\section{Conclusion}
We have demonstrated that the Normalizing Flows–based framework, previously shown to achieve high-quality reconstruction of spin-based observables in $t/\bar{t} + DM$ mediator topologies, can be successfully extended to the reconstruction of spin correlations in dileptonic $t\bar t$ events. By learning the full conditional probability density, both the basic autoregressive flows and the more expressive $\nu$-Flows preserve multi-dimensional correlations and deliver high-fidelity recovery of the entanglement marker $\cos\varphi$ across distinct $m_{t\bar t}$ regions. While a simple MLP can achieve competitive quality in this final state, it falls behind in single-top production and is less suitable for the unified approach to reconstruct spin correlations. 

Looking ahead, this paradigm naturally generalizes to rare multi–top processes such as $t\bar t tW$ and $t\bar t t\bar t$, where the complexity of invisible-momentum reconstruction and combinatorial assignment is greatly amplified. The development of unified architectures—for instance, combining flows with transformer–based unfolding and resonance matching—promises to enable precision studies of spin correlations and enhance sensitivity to dark matter in three– and four–top final states.

\section*{Acknowledgments}
\label{sec:acknowledgement}
This study was conducted within the scientific program of the Russian National Center for Physics and Mathematics, section 5 «Particle Physics and Cosmology».
\clearpage

\end{document}